# Search for Terrestrial Intelligence

Olivier Auber.
Evolution, Complexity and COgnition (ECCO lab) at Vrije Universiteit Brussel (VUB).
Global Brain Institute. P2P foundation.

**The existence of intelligent extraterrestrial life forms among the thousands of billions of galaxies that dance in the Universe is beyond doubt (1). Meeting with one of them is only a matter of time. SETI (Search for Extraterrestrial Intelligence) has been searching for them for only half a century. However, the crucial question today is whether we will have time to establish contact given the multiple threats that humanity puts on itself. Therefore, it seems urgent to launch a program complementary to SETI which could be called STI (Search for Terrestrial Intelligence).**

In 1950, the physicist Enrico Fermi formulated the following paradox: *If there were extraterrestrial civilizations, their representatives should be already around us. But we do not see them.* One of the hypotheses that could explain their absence could be that intelligent civilizations all end up self-destructing before making any contact with one another. A mysterious phenomenon would be at work at the heart of civilizations; Suicidal behavior would emerge systematically from all forms of intelligent lives.

Concerning humanity, some evoke a *systemic stupidity* (2) specific to our species. On the scale of the Universe, if this hypothesis is true, one could speak of *cosmic stupidity*.

All these issues have been debated since the beginnings of SETI. *Cosmic stupidity* is even put into equations. However, it was not SETI's vocation to tackle this problem directly. To do so today with a STI program, one should start by taking seriously a hypothesis contrary to that of *cosmic stupidity*.

**Existential Hypothesis | STI H1:** *There is a non-zero chance that some advanced civilizations will successfully overcome their propensity for self-destruction.*

Certainly, the absence of our extraterrestrial contacts seems to corroborate the scarcity of this evolutionary performance, but to date, nothing in the observations of SETI prohibits the exclusion of this hypothesis. Obviously, everyone is entitled to express reservations when he/she looks at the state of the planet Earth. So how could a STI program be able to escape our *systemic stupidity*? At least how could we give SETI enough time to spot a civilization that refutes *cosmic stupidity*? Who knows, it might be able to teach us how to fight our own *stupidity*?

In the same way as the SETI@home initiative, which allows everyone to contribute to the decoding of extraterrestrial signals from home, an **STI@home** initiative could propose to participate in the creation of a network of detectors and amplifiers of human intelligence, capable of revealing the mechanisms of our *stupidity*, and of neutralizing them. It goes without saying that this should be done in a very special way so as not to once again put human intelligence at the service of *stupidity*. The already quoted Enrico Fermi, one of the fathers of the atomic bomb, knows something about it.

Let us try first to situate the magnitude of the challenge and to formulate some hypotheses that would support this idea. Undoubtedly, STI would ultimately involve solving the most arduous scientific questions (hard problems), such as those of the origin of the universe, life, language and consciousness. All this of course, in a sense that does not turn against humanity. But in a start-up phase, STI could focus on a more limited problem directly related to the paradox of our *systemic stupidity*.

This paradox can be expressed as follows: Global crises (ecological, economic, financial, political, cultural, religious), ecocides, genocides and wars that we create from scratch, are network phenomena that far exceed their individual participants, who are intelligent, even very intelligent for some, and not suicidal for the most part.

At the heart of the paradox, there are networks and what we do with them. STI could be limited to this at first and invite us to ask ourselves the following questions: What status do we give to our networks? What information dynamics animate them? What evolution do we envision for our artificial networks in relation to our biological, social and cultural nature?

Let us note that the ambition of STI, limited to these questions relating to artificial networks already implies the resolution of many conundrums. In particular, there is no model for predicting the dynamics of a network in which all agents try to predict the predictions of all others. For example, nobody holds a Great Model to predict the evolution of financial markets in which all actors act with their own predictive models. We just observe that markets form bubbles and repetitive crises always more catastrophic than the previous ones, without knowing why or how.

And there is a more fundamental problem. Assuming that such a Great Model exists one day, how could it predict its own influence on the dynamics and evolution of networks once it is recognized and applied by all its agents? This question of the feedback of the model on itself leads to an aporia to which are confronted all the attempts of modeling such as the Global Brain, which many scientists have been working on for years. To try to solve this paradox and the mysteries that are linked to it, let us try some hypotheses.

***Topological hypothesis | STI H2:*** *All intelligent life forms have artificial Internet-like networks. These networks are developed according to certain particular topologies rather than as random simplicial complexes. That is to say that extraterrestrials communicate with each other, as we do ourselves, via centralized networks and/or distributed or meshed networks, quantum in their extreme sophistication. These networks are the backbone of their society just as they are the backbone of ours.*

***Evolutionary Hypothesis | STI H3:*** *There are universal laws for the implementation of networks and their artifacts that provide the species that deploy them with an evolutionary advantage. These laws are mathematical. Civilizations discover them little by little by variation, mutation, selection. The few civilizations that apply these evolutionary laws in time are led to survive while the vast majority is brought to extinction.*

Presumably, before the laws of the networks are elucidated and applied (or not), civilizations would grope around and elaborate all sorts of more or less relevant and adapted theories and practices, which would ultimately be subjected to natural selection. Let's examine three scenarios among infinite possibilities. Let's start with the most extreme.

### *Devoratus*-style civilizations

Some intelligent species could follow the propensity of the human species to the predation of all available resources and to the constant rise of energy consumption. A vertiginous extrapolation (3) leads to imagine civilizations gathered around stars from which they would capture all or part of their energy. It is the hypothesis of the Dyson Spheres (Dyson, 1960) of which SETI thought it had detected an example not long ago. By extrapolating this trend, one can imagine stellivorous life forms that literally devour the stars (Vidal, 2016). Could SETI also have found a trace of it? Why not? We could also imagine civilizations entirely confused with black holes, dark matter, or multiverses. The intelligent entities within these civilizations are unlikely to have anything biological. Their artificial networks would no longer have anything material in the sense in which we usually understand it. Everything would be fused in a kind of quantum maelstrom. One can interpret this fusion as the ultimate evolution of civilizations collapsed on themselves under the weight of *cosmic stupidity*. It can also be interpreted in a different sense. Indeed, it is not at all obvious that there is a continuity between life as we think we know it and these intelligent forms with very high energy. These could be two fully independent branches of life, although they may share common patterns. Be that as it may, the fact that high energy life forms are the first to be guessed is not surprising. The *devoratus*-style civilizations are indeed the only ones capable of forming signals powerful enough to reach us. This does not prevent us from imagining that other civilizations with lower energy, and therefore less detectable but closer to us, cannot thrive on the basis of STI assumptions.

### *Formabilis*-style civilizations

Some civilizations could imagine as certain humans do, that their evolution would be guided by *morphic resonance* (Sheldrake, 1981). This theory is supposed to explain some phenomena of synchrony and precognition, especially in networks where all agents predict the behavior of others, but also the replications between living systems without apparent physical links such as distant civilizations. To test this hypothesis, as already proposed by some humans (Goertzel, 2016), a civilization could build a General Artificial Intelligence (AGI) which would far exceed the cognitive capacities of its population. This AGI would be supposed capable of capturing realities inaccessible to those who created it. Notably, it could discover new mathematical laws, supposed to be the type of information most easily transmitted from civilizations to civilizations thanks to *morphic fields*. According to the evolutionary hypothesis STI H2, a civilization of this style would survive only if its AGI succeeds in discovering the evolutionary laws of networks and applying them to itself. This is unlikely to happen: a) if the theory of *morphic resonance* is false. It would indeed constitute a bias preventing the AGI from calculating coherent laws; b) if the theory of *morphic resonance* is true and if the hypothesis of *cosmic stupidity* is true too. In this case, the AGI could only end up with its own suicide and that of the *formabilis* civilization that created it.

On the other hand, the AGI could possibly confirm the evolutionary hypothesis STI H2 and neutralize its *systemic stupidity* if the existential hypothesis STI H1 and the theory of the *morphic resonance* are both true. Let us note that the chances are low because today the theory in question is far from being endorsed by the majority of scientists of planet Earth.

### *Imitativus*-stlyle civilizations

Other civilizations facing as we do the mystery of their evolution, could imagine that they are basically the toys of invisible informational patterns, which would use them to evolve and replicate. According to this belief, these patterns would survive beyond the civilizations that convey them, in forms that would evade the senses and the understanding of intelligent beings. These beings would be in fact only sorts of computer simulations driven by some quantum demiurge. This somewhat animistic hypothesis has a more scientific version on planet Earth. It is known as the *memetics theory.* The patterns in question are called *memes*: contracted term of imitation and gene (Dawkins, 1976). As some humans propose, it is possible to admit that "*memes which succeed in inserting themselves in the edifice of representations are those which allow an optimization of the topology of the system and its access to resources […]. Those that are rejected are those that impose detours and delays to the optimized architecture or to the functioning of the system.*" (Baquiast, Jacquemin, 2003). *Memetics* thus appears as a sub-thesis of the theory of evolution, but does not give *a priori* any means to guide it. Consider a civilization that would remain there. Practically, in the hopelessness of being able to control its long-term evolution, its social forces would try by all means to influence and manipulate *memes* in a way that is favorable to them in the short term. Its networks would be transformed into *memetic* battlefields. In fact, a kind of invisible cyber warfare would reinforce the *systemic stupidity*. According to the evolutionary hypothesis STI H2, *Imitativus*-style civilizations would therefore have little chance of discovering the evolutionary laws of their networks and their survival would be limited.

*Devoratus*, *formabilis* and *imitativus*-style civilizations, described as being able to develop in distant galaxies, could of course evolve in the future of the planet Earth. The evolutionary hypothesis STI H2 gives an indication of our low chances of survival in these cases. Let us see, using other hypotheses, how STI could help to identify the nature of the evolutionary laws of networks and thus open a path towards a positive and lasting outcome for our species. In this sense, a good track to follow is that of information; the raw material of networks.

Seventy years of research devoted to the elaboration of a theory of information have led to the conclusion that information is not an absolute value independent of the context but rather a relative one. The nature of the information can be summarized as follows:

- Information is a drop in complexity (unexpected, difficult to obtain, simple to describe).
- Information only exists if it can be "read".
- Information is what "survives".

On this basis, STI could formulate a new hypothesis.

**Hypothesis of the universality of information | STI H4:** *The nature of information is universal. It is valid on Earth as in any galaxy. Here, as elsewhere, all the intelligent entities composing civilizations are thus linked together and with their environment by the thread of information.*

A somewhat counter-intuitive consequence, already mentioned here concerning human beings, is the following. Whether we like it or not, we compete, not just for resources and for reproduction, but essentially *to provide information to others.* From this competition emerges a form of social hierarchy revealing three classes of individuals distinguishable according to their attractiveness (conditioning the size of their social networks), and their signalling activity (the quantity of messages they emit). The competition between individuals is weak for the two extreme classes (the one with a small social network and the one with a large social network). Competition is much stronger for the intermediate class of individuals with an average social network. This class distribution is an Evolutionarily Stable Strategy (ESS) of our species, linked to the origin and functions of our language (Dessalles, 2016). It should be noted that this social structure, however automatic it may be, seems less deterministic than that of other species, for example that of tilapia fish (4), because humans are supposed to have the cognitive capacity to self-reflect on their own social structure.

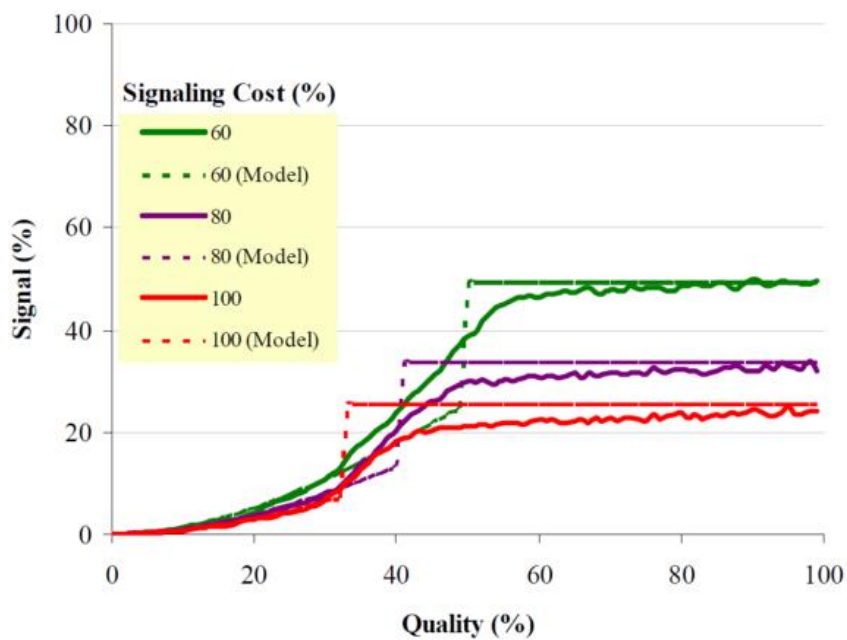

**Figure 1 : Dessalles, J-L. (2014). Optimal Investment in Social Signals.**

If we follow the universality of information STI H4 hypothesis, it is very likely that some intelligent extraterrestrial entities with cognitive equipment comparable to ours reproduce this same social scheme. Result: There would also be social classes elsewhere in the universe. Class struggles too!

At this stage, we see that here as elsewhere, intelligence has an ambivalent position in the emergence of *systemic stupidity*. It is, on the one hand, an instrument often used for the benefit of domination which, as we have just seen, is practiced in a quasi-mechanical way.

On the other hand, and in the best of cases, intelligence is an intellectual tool that can allow to observe its own mechanisms and its own *stupidity*. This is obviously what STI is looking for.

The STI H4 hypothesis, complementing the three preceding scenarios, leads us to imagine a fourth woesome scenario and a fifth one offering some avenues to avoid extinction.

*Pervasus*-**style civilizations.**

It is not difficult to imagine advanced civilizations having developed their artificial networks to the point of making them entirely pervasive, that is, connected to the least vital and cognitive function of all members of their population. It is not difficult to imagine also that these networks are controlled exclusively by a very small dominant class, which would ensure it introduces all the biases necessary to maintain its dominance in the algorithms regulating the behavior of the network and each of its agents. A predictable consequence is that the class structure of such an extraterrestrial civilization would be totally crystallized. In other words, this would lead to a type of society that the sociologist Niklas Luhmann analyzes as a particular case of *autopoietic* system (self-referential), based on a single *dichotomous code*: inside/outside (Luhmann, 2011). The characteristic of such systems is their extreme fragility in the face of changes, due to lack of capacity for variation and mutation. It is therefore probable that a *pervasus*-style civilization would disappear fairly quickly on the occasion of various environmental shocks or internal disturbances which would most likely happen.

*Legitimus*-**style civilizations.**

Let us now imagine the evolution of highly developed extraterrestrial civilizations that would have initiated large-scale STI programs in time. Within a few decades or a few centuries, all the intelligent entities composing a civilization of this type would have integrated the four STI hypotheses. In particular, they would have understood the nature of the information linking them, its consequences in terms of social structure, the influence of network topology and, therefore, the dangers presented by the previous scenarios. They would also have done all kinds of experiments on their networks to try to understand their dynamics. In particular, they would have practiced massively a contemplative game such as the Poietic Generator (known on Earth in an embryonic form) which would have helped them to take a step back on the functioning of the networks. This would have given them the idea of adding two new assumptions to their STI program to clarify the evolutionary laws of their networks.

**Hypothesis of the anoptical perspectives | STI H5:** *All networks have a certain shape. Some shapes simpler than others, offer a complexity drop for their agents. The simplest*

networks have a "centralized" shape. Their central point operates on the network in the manner of a "vanishing point" of a perspective that is not spatial but temporal. It is at the temporal vanishing point that agents' collective time emerges out of the network. Other networks are simply "distributed" or "meshed". They do not have a physical vanishing point like centralized networks, but it is an arbitrary code that plays this role. It is simply the sign of recognition under cover of which agents exchange. This code operates on the network in the manner of a vanishing code of a perspective that is no longer only temporal but digital. These two perspectives (temporal and digital) are called anoptical because of the type of "reading" of the shape of the network that they offer to agents. Indeed this "reading" is not optical but it implies all their cognitive abilities.

**Hypothesis of legitimacy | STI H6:** In the same way that there exists a "legitimate construction" of the optical perspective based on geometrical principles, there exists a "legitimate construction" of anoptical perspectives based on cognitive principles. The legitimacy of networks' construction is an evolutionary advantage. The proposed criteria of legitimacy (5) are the following:

> **A)** Does any agent A have the actual right to access the network if he requests it? Can A leave the network freely?
>
> **AB)** Is any agent B (present or future, including agents that conceive, administer, and develop the network) treated like A?
>
> **ABC)** If agents A, B, and C (where ABC is the beginning of a multitude) belong to a network that meets the first two criteria, are they peers? That is, are they able to recognize, trust, and respect each other, thereby building common representations and common sense?

At this stage, these last two assumptions STI H5 and STI H6 have no force of law, either in a mathematical sense or in a legal sense. Nevertheless, *legitimus*-style civilizations would have decided to apply them on a large scale on their networks, thus making them *legitimate*. Somehow, they would have found a way to thwart much of their *systemic stupidity* and the feeling of progressing towards a certain mastery of their evolution.

**Returning to Earth**

It is easy to imagine that a ***legitimus***-style civilization can evolve in a distant galaxy. It is much more difficult to imagine that it can emerge here where intelligence seems too often trapped in *systemic stupidity*.

- On Earth as in *devoratus* civilizations, the trend remains predation of resources and increase in energy consumption, with the ecological and geopolitical consequences that we know. The networks activated in this game (energetical, financial, political, military) are clearly not *legitimate* within the meaning of the STI H6 criteria.

- As in *formabilis* civilizations, most of the efforts of research and industry converge towards the development of some sort of Artificial Intelligence pretending to surpass that of the population. Humans have the right to ask themselves what their place is in the midst of AIs, and therefore to question the *legitimacy* of the networks on which they are proliferating.

- As in *imitativus* civilizations, terrestrial networks are a battleground for the control of opinions, emotions, behaviors and personal data. Everything happens as if some actors were taking advantage of the humans' confinement in cognitive bubbles and of their inability to claim the *legitimacy* of the networks that maintain them confined.

- As in *pervasus* civilizations, terrestrial technologies held by an ever smaller number of individuals claim to regulate all the vital and cognitive functions of all the others. Natural competition between individuals is increasingly engraved in silicon and soon in manipulated genes. Social classes threaten to turn into castes, the most favored of which is dreaming to become immortal. The breakdown of *legitimacy* would thus be definitive.

The situation seems desperate. Yet examples in human history show how a change of representation can spread very quickly in all minds and overthrow a society that is supposed to be immutable. This is particularly the case with the invention of printing press (Gutenberg, 1448) and that of optical perspective (Brunelleschi, 1413). Within a few centuries, with optical perspective, which, it should be pointed out, is not even a technology but a mere operation of mind, all the hierarchical edifice of the Middle Ages made way for the geometric construction of the Renaissance. In the end, it is the very place of human as being in the world that has been redefined.

Of course, a kind of STI program already exists on Earth. Billions of researchers have been busy for millions of years to find the forms of intelligence that could free us from our *systemic stupidity*. The invention of optical perspective was one of the major breakthroughs of our research. Six centuries later, if the results are so disappointing to the point that this *stupidity* threatens to prevail, may it be because we need another breakthrough?

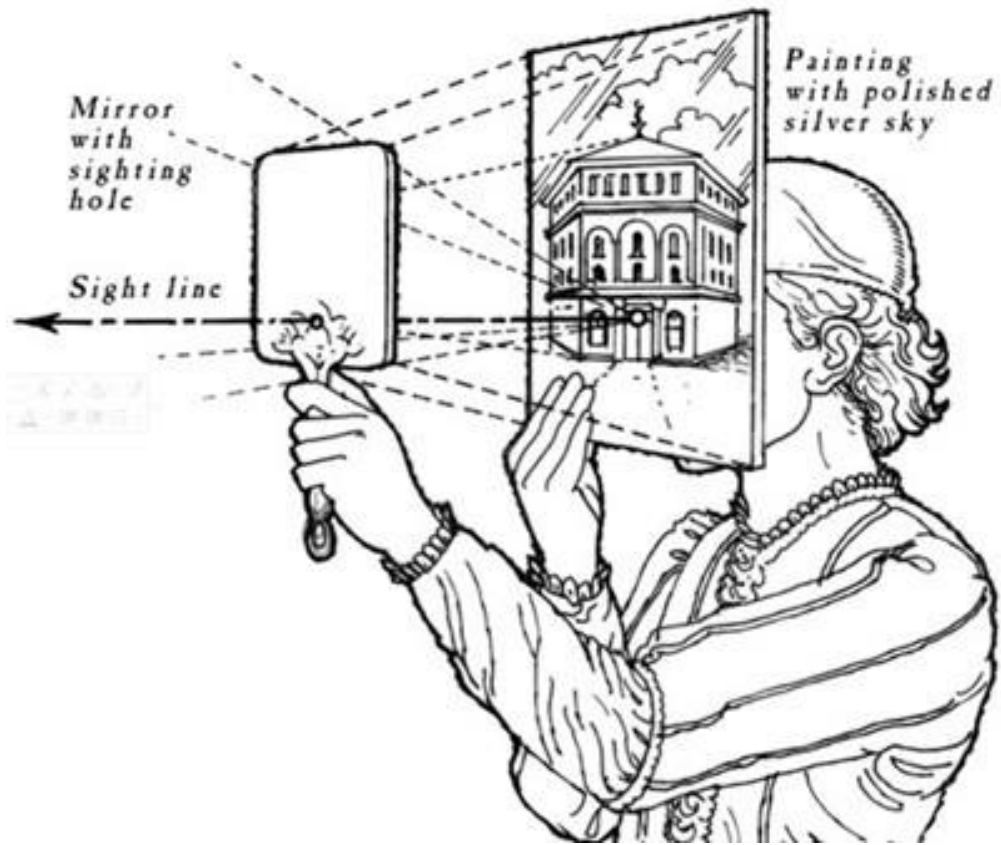

**Filippo Brunelleschi (1377–April 15, 1446). Two panel paintings illustrating for the first time around 1413 the geometry of the optical perspective.**

The optical perspective of the Renaissance has established us in our capacity to all be observers and actors of a geometric construction of the visible world. However, it had to be admitted that the *vanishing point* of this construction—symbol of the infinite and the unknowable—escapes the terrestrial political and religious order. Today, we understand little by little that we are part of invisible networks, interweaving the natural with the artificial at all scales. It is not too difficult to admit that these networks are all coupled to one another and form an unique and immense network, which one will doubtless never be able to grasp in thought. But the problem is that we have not yet realized that we construct the artificial part of this network by using an *anoptical perspective* that implies all of our cognition. We have not yet assimilated the criteria of *legitimacy* of this construction, nor have we admitted that its ultimate *vanishing code* (6) establishing the connection between all networks and their co-evolution, is also a symbol of the infinite and the unknowable that escapes the earthly order.

However, small causes could have great effects. A change of representation could emerge. It only misses a click to realize that networks offer us a game that plays on Earth at the scale of the universe. The challenge? Nothing less than finding forms of intelligence that can refute *cosmic stupidity*. Everyone has a role to play.

*A very modest embryo of STI@home initiative is here. Some avenues to understand the dynamics of networks may be found in this science paper (PDF). Many other initiatives and research are invited to complete it*

***Notes:***
*(1) Provisional estimate (2016).*
*(2) In the words of Bernard Stiegler: "bêtise systémique".*
*(3) This extrapolation was made for the first time in 1960 by the Soviet astronomer Nikolai Kardachev, which led him to define three types of civilizations according to their energy consumption. They are all included here in the devoratus-style.*
*(4) The social and sexual behavior of the dominant male of tilapia fish depends on a gene that also gives it a distinctive color and hence the recognition of its status by the group. But this gene is controlled by the social environment: it is inhibited in the dominant by the introduction into the aquarium of a larger male, and activated in the subordinate when the dominant is removed (Source).*
*(5) These criteria are detailed in the article available on Noemalab: Refounding Legitimacy, Towards Aethogenesis.*
*(6) Metric? Field? Motive? Information principle?*